\documentclass[a4paper,11pt]{article}

\usepackage{graphicx}

\usepackage[all]{xy}
\usepackage{latexsym,amssymb,amsmath,amsfonts, amsthm, xcolor}

\usepackage[all]{xy}
\usepackage{latexsym,amssymb,amsmath,amsfonts, amsthm, xcolor}

\def\vbar{\mathchoice{\vrule height6.3ptdepth-.5ptwidth.8pt\kern-.8pt}
   {\vrule height6.3ptdepth-.5ptwidth.8pt\kern-.8pt}
   {\vrule height4.1ptdepth-.35ptwidth.6pt\kern-.6pt}
   {\vrule height3.1ptdepth-.25ptwidth.5pt\kern-.5pt}}
\def\fudge{\mathchoice{}{}{\mkern.5mu}{\mkern.8mu}}
\def\bbc#1#2{{\rm \mkern#2mu\vbar\mkern-#2mu#1}}
\def\bbb#1{{\rm I\mkern-3.5mu #1}}
\def\bba#1#2{{\rm #1\mkern-#2mu\fudge #1}}
\def\bb#1{{\count4=`#1 \advance\count4by-64 \ifcase\count4\or\bba A{11.5}\or
   \bbb B\or\bbc C{5}\or\bbb D\or\bbb E\or\bbb F \or\bbc G{5}\or\bbb H\or
   \bbb I\or\bbc J{3}\or\bbb K\or\bbb L \or\bbb M\or\bbb N\or\bbc O{5} \or
   \bbb P\or\bbc Q{5}\or\bbb R\or\bbc S{4.2}\or\bba T{10.5}\or\bbc U{5}\or
   \bba V{12}\or\bba W{16.5}\or\bba X{11}\or\bba Y{11.7}\or\bba Z{7.5}\fi}}

\newcommand{\RR}{\mbox{${\rm \:  R\!\!\!\! I
\;\;}$}}

\newcommand{\vs}{\vspace{0.25cm}}

\newtheorem{theorem}{Theorem}
\newtheorem{itlemma}{Lemma}[section]
\newtheorem{itproposition}[itlemma]{Proposition}
\newtheorem{itcorollary}[itlemma]{Corollary}
\newtheorem{itremark}[itlemma]{Remark}
\newtheorem{itremarks}[itlemma]{Remarks}
\newtheorem{itdefinition}[itlemma]{Definition}
\newtheorem{itexample}[itlemma]{Example}

\newenvironment{lemma}{\begin{itlemma}\rm}{\end{itlemma}} 
\newenvironment{remark}{\begin{itremark}\rm}{\end{itremark}} 
\newenvironment{remarks}{\begin{itremarks} \rm}{\end{itremarks}}
\newenvironment{corollary}{\begin{itcorollary}\rm}{\end{itcorollary}}
\newenvironment{proposition}{\begin{itproposition}\rm}{\end{itproposition}}
\newenvironment{definition}{\begin{itdefinition}\rm}{\end{itdefinition}}
\newenvironment{example}{\begin{itexample}\rm}{\end{itexample}}
\newenvironment{fact}{\noindent {\em Fact}. \ \ }{\hfill \medskip}
\newenvironment{claim}{\noindent {\em Claim}. \ \ }{\hfill \medskip}
\newcommand{\be}[1]{\begin{equation}\label{#1}}
\newcommand{\ee}{\end{equation}}
\newcommand{\bl}[1]{\begin{lemma}\label{#1}}
\newcommand{\br}[1]{\begin{remark}\label{#1}}
\newcommand{\brs}[1]{\begin{remarks}\label{#1}}
\newcommand{\bt}[1]{\begin{theorem}\label{#1}}
\newcommand{\bd}[1]{\begin{definition}\label{#1}}
\newcommand{\bp}[1]{\begin{proposition}\label{#1}}
\newcommand{\bc}[1]{\begin{corollary}\label{#1}}
\newcommand{\bfact}[1]{\begin{fact}\label{#1}}
\newcommand{\bex}[1]{\begin{example}\label{#1}}
\newcommand{\ec}{\end{corollary}}
\newcommand{\efact}{\end{fact}}
\newcommand{\eex}{\end{example}}
\newcommand{\el}{\end{lemma}}
\newcommand{\er}{\end{remark}}
\newcommand{\ers}{\end{remarks}}
\newcommand{\et}{\end{theorem}}
\newcommand{\ed}{\end{definition}}
\newcommand{\ep}{\end{proposition}}
\newcommand{\epr}{\end{proof}}
\newcommand{\bpr}{\begin{proof}}
\newcommand{\bcl}{\begin{claim}}
\newcommand{\ecl}{\end{claim}}

\newcommand{\bi}{\begin{itemize}}
\newcommand{\ei}{\end{itemize}}
\newcommand{\ben}{\begin{enumerate}}
\newcommand{\een}{\end{enumerate}}

\begin{document}

\begin{center}

{\Large{Equivalence Between Indirect Controllability and Complete
Controllability for Quantum Systems}}

\vs


{\large{Domenico D'Alessandro \\ Department of Mathematics, Iowa
State
 University, \\  Ames IA-5001, Iowa, U.S.A.;\\
Electronic address: dmdaless@gmail.com.}}

\end{center}

\vs

\vs

\begin{center}

{July 27, 2012}

\end{center}
\vs

\begin{abstract} We consider a control scheme where a quantum system
$S$ is put in contact with an auxiliary quantum system $A$ and the
control can affect $A$ only, while  $S$ is the system of interest.
The system $S$ is then controlled {\it indirectly} through the
interaction with $A$. {\it Complete controllability} of $S+A$  means
that every unitary state transformation for  the system $S+A$ can be
achieved with this scheme. {\it Indirect controllability} means that
every unitary transformation on the system $S$ can be achieved. We
prove in this paper, under appropriate conditions and definitions,
that these two notions are equivalent in finite dimension. We use
Lie algebraic methods to prove this result.

\end{abstract}

\vspace{0.5cm}


\section{Introduction}

In many experimental set-ups,  a quantum system $S$, which is the
{\it target} of control, is put in contact with an {\it auxiliary}
quantum system $A$ and the control  can only directly affect $A$,
while $S$ is the system of interest. Therefore $S$ is controlled
{\it indirectly} via the interaction with $A$. The ({\it indirect})
{\it controllability} of $S$ with this scheme has been studied in
several papers and for  various physical examples   (see, e.g.,
\cite{ind1}, \cite{ind2}). However always conditions have been given
so that the full system $S+A$ is completely controllable, i.e.,
every unitary transformation can be achieved in the Hilbert space
associated with the full system. This implies in particular that $S$
is indirectly controllable, i.e., any unitary transformation on the
state of $S$ can be obtained. The opposite is in general not true
and there are schemes  where one can have indirect controllability
of the system $S$ without having complete controllability of the
full system $S+A$. An example of this was given in \cite{IoeRaf}
(Proposition 5.2)  for the case of two coupled qubits $S$ and $A$.
Whether or not we can have indirect controllability of $S$ without
complete controllability of $S+A$, depends in general on the initial
state assumed for $A$. In this paper we shall prove that if the
initial state of $A$ is the perfectly mixed state (see definitions
below), then complete controllability is also necessary to have
indirect controllability. Therefore if we require indirect
controllability for an {\it arbitrary state} of the system $A$ the
two definitions are equivalent. We now describe in mathematical
terms  the definitions and result of this paper.


The state of a finite dimensional quantum mechanical system is
represented by a {\it density matrix}, that is, a trace $1$,
positive semidefinite Hermitian matrix acting as a linear operator
on a Hilbert space associated with the system. The {\it dimension}
of the system refers to the dimension of this Hilbert space. We
shall denote by $\rho_S$, $\rho_A$, and $\rho_{TOT}$, the density
matrices for the systems $S$, $A$ and $S+A$, respectively, which
have dimensions $n_S$, $n_A$ and $n_S n_A$, respectively. The
density matrix $\rho_S$ ($\rho_A$) is obtained from $\rho_{TOT}$
through  the operation of {\it partial trace} with respect to $A$
($S$), that is \be{parttrace} \rho_S=Tr_A(\rho_{TOT}), \qquad
\rho_A=Tr_S(\rho_{TOT}).\ee The dynamics of the total system is
determined by \be{dyntot} \rho_{TOT}(t)=X_{TOT}(t) \rho_{TOT}(0)
X^{\dagger}_{TOT}(t),  \ee where $X_{TOT}$ is the solution of {\it
Schr\"odinger operator equation} \be{Scropeq} i\dot X_{TOT}:= H(u)
X_{TOT}, \qquad X_{TOT}(0)={\bf 1}_{n_Sn_A}. \ee In (\ref{Scropeq}),
${\bf 1}_{n_Sn_A}$ is the $n_Sn_A \times n_S n_A$
identity\footnote{In the following, ${\bf 1}_v$ denotes the $v
\times v$ identity. We shall omit the index $v$ when the dimension
is obvious from the context.} and $H(u)$ is the {\it Hamiltonian
operator}, an $n_S n_A \times n_S n_A$ Hermitian matrix which we
assume function of a {\it control} $u$.

According to the {\it Lie algebra rank condition} \cite{JS} applied
to quantum control (see, e.g., \cite{mikobook}), the set ${\cal R}$
of possible transformations, $X_{TOT}$, which can be obtained as
solutions of (\ref{Scropeq}) is as follows. Let ${\cal L}$ be the
Lie algebra generated by the set \be{Liaalggenerat} {\cal F}:=\{
iH(u) \, | \, u \in {\cal U}\},  \ee where ${\cal U}$ is the set of
possible values for the control $u$. Denote by $e^{\cal L}$ the
associated Lie group. If $e^{\cal L}$ is compact, then ${\cal R}$ is
equal to $e^{\cal L}$. If $e^{\cal L}$ is not compact, then ${\cal
R}$ is dense in $e^{\cal L}$.\footnote{This last statement is a
consequence of the fact that for quantum systems $e^{\cal L}$ is
always a subgroup of the unitary Lie group $U(n_S n_A)$ (cf.
\cite{mikocool}, \cite{shapiro}).} In the following, in order not to
complicate the exposition, we shall neglect this distinction and
always assume ${\cal R}=e^{\cal L}$ where sometimes the equality
between two topological spaces really means that one space is dense
in the other. The Lie algebra ${\cal L}$ is called the {\it
dynamical Lie algebra} associated with the system $S+A$. If ${\cal
L}$ is the full $u(n_S n_A)$ or $su(n_S n_A)$,\footnote{The Lie
algebras of $n_S n_A \times n_S n_A$ skew-Hermitian matrices or $n_S
n_A \times n_S n_A$ skew-Hermitian matrices with zero trace,
respectively.} then the system $S+A$ is called {\it completely
controllable} and ${\cal R}$ is $U(n_S n_A)$ or $SU(n_S n_A)$,
respectively.\footnote{The full Lie group  of $n_S n_A \times n_S
n_A$ unitary matrices or the full Lie group  of $n_S n_A \times n_S
n_A$ unitary matrices with determinant equal to one, respectively.}
In this case, every unitary transformation on the initial state
$\rho_{TOT}(0)$ according to (\ref{dyntot}) is possible.

We shall assume in this paper that systems $S$ and $A$ are initially
uncorrelated, i.e., the initial state $\rho_{TOT}(0)$ has the form
$\rho_{TOT}(0)=\rho_S(0) \otimes \rho_{A}(0)$. The evolution of the
target system $S$ is obtained by combining  (\ref{dyntot}) with
(\ref{parttrace}), i.e., \be{combinedevo} \rho_S(t)=Tr_A
\left(X_{TOT}(t) \rho_S(0) \otimes \rho_A(0) X_{TOT}^\dagger (t)
\right), \ee where $X_{TOT}$ is the solution of (\ref{Scropeq}).
Therefore,  the set of available states for the system $S$, starting
from $\rho_S$, and with $A$ in the initial state $\rho_A$,  is
\be{availablestates} {\cal R}_S:=\left\{Tr_A(X_{TOT} \rho_S \otimes
\rho_A X^\dagger_{TOT})| X_{TOT} \in e^{\cal L}\right\}. \ee In
indirect control schemes, the set of generators of the dynamical Lie
algebra ${\cal L}$, i.e., ${\cal F}$ in (\ref{Liaalggenerat}),  is
to  be taken of the form \be{Liealggenindirect} {\cal F}:=\{ J \}
\cup \{ \tilde  {\cal B}\}, \ee where the set $\tilde {\cal B}$
generates a  Lie subalgebra ${\cal B}$ of $u(n_Sn_A)$ of matrices of
the form ${\bf 1}_{n_S} \otimes B$ with $B$ in $u(n_A)$. This
subalgebra describes the {\it control authority} we have on the
auxiliary system $A$. Transformations in the associated Lie group,
$e^{\cal B}$, are all available and they are of the form ${\bf 1}
\otimes X_A$,  with $X_A \in U(n_A)$. Therefore any initial state
$\rho_S \otimes \rho_A$ can be transformed as \be{transformat23}
\rho_S \otimes \rho_A \rightarrow ({\bf 1} \otimes X_A) \rho_S
\otimes \rho_A  ({\bf 1} \otimes X_A^\dagger)= \rho_S \otimes (X_A
\rho_A X_A^\dagger). \ee In (\ref{Liealggenindirect}) The
(Hamiltonian) matrix $J$ models the autonomous (non-controlled)
dynamics of the system $S$, the autonomous dynamics of the system
$A$ and the {\it interaction} between the system $S$ and the
auxiliary system $A$. These three terms, in that order,  are the
three summands in the definition of $J$ \be{Jform} J:=K \otimes {\bf
1} + {\bf 1} \otimes L + \sum_{j=1}^n iS_j \otimes \sigma_j. \ee
 Here $K$ and $S_j$, $j=1,\ldots,n$, are in $su(n_S)$, $L$ and $\sigma_j$,
 $j=1,\ldots,n$,
 are in $su(n_A)$ and  the $\sigma_j$'s are linearly independent. In the
 following, we shall assume that ${\cal B}$ does not contain any
 nonzero trace element, so that the dynamical Lie algebra ${\cal
 L}$ is a Lie subalgebra of $su(n_S n_A)$. This is done
 without loss of generality as multiples of the identity only induce
 a common phase factor in equation (\ref{Scropeq}) which has no effect on the
 dynamics of $\rho_{TOT}$ in (\ref{dyntot}).

 There are several notions of indirect controllability \cite{IoeRaf}, according to the
 restrictions we place  on the possible initial states for the auxiliary
 system $A$
 and the possible states we require to reach for the system $S$, starting from $\rho_S(0)$
 (e.g., unitary equivalent, or general density matrices). We shall adopt, in this
 paper, the following definitions (cf. \cite{IoeRaf}).

\bd{indcongivenstate} The system $S$ is called {\it indirectly
controllable} given $\rho_A$ (initial state of $A$) if,
 for every initial density matrix  $\rho_S$  and every unitary $X_S \in
 U(n_S)$,
 there exists a (reachable)  $X_{TOT} \in e^{\cal L}$ such that (cf. (\ref{combinedevo}))
 \be{unita}
 X_S \rho_S X_S^\dagger=Tr_A(X_{TOT} \rho_S \otimes \rho_A X_{TOT}^\dagger).
 \ee
\ed

 \bd{stronglyinUnit}
 The system $S$ is called {\it strongly indirectly controllable} if it
 is {\it
indirectly controllable} given $\rho_A$ for every initial state
$\rho_A$ of $A$. \ed In other terms, we are able to steer the state
of the system $S$ between any two unitarily equivalent states
independently of the state $\rho_A$ of the auxiliary system $A$. The
indirect control scheme works just as well as a completely
controllable scheme for system $S$. It was proven in \cite{IoeRaf},
for the case where both $S$ and $A$ are qubits, and every unitary is
available on the system $A$ (i.e., ${\cal B}=su(n_A)$ above), that
this property is equivalent to complete controllability of the total
system. The goal of this paper is to extend this result to the case
where $S$ and $A$ have arbitrary dimensions. In particular, our main
result is as follows:

\bt{MainT} Assume ${\cal B}= su(n_A)$. A system $S$ is indirectly
controllable given the perfectly mixed state $\rho_A:=\frac{1}{n_A}
{\bf 1}$ for $A$ if and only if the total system $S+A$ is completely
controllable. Therefore it is strongly indirectly controllable if
and only if the system $S+A$ is completely controllable. \et

Indirect controllability can be studied using Lie algebraic methods
but the investigation is complicated by the fact that the various
controllability notions are not invariant under general (unitary)
coordinate transformations in the state space of the system $S+A$.
They are invariant  only under  {\it local} transformations, that
is, transformations which act on the Hilbert spaces of $S$ and $A$
separately. For instance, if we replace the dynamical Lie algebra
$\cal L$ with ${\cal L}':=(T_S \otimes T_A) {\cal L} (T_S^\dagger
\otimes T_A^\dagger)$, with $T_S \in U(n_S)$ and $T_A \in U(n_A)$,
then indirect controllability is not modified as it can be easily
seen using the property of the partial trace \be{propparttra} Tr_A
((T_S \otimes T_A) \rho_{TOT} (T_S^\dagger \otimes T_A^\dagger))=T_S
Tr_A(\rho_{TOT}) T_S^\dagger. \ee One direction of Theorem
\ref{MainT} follows immediately from the property
(\ref{propparttra}) of the partial trace. In fact, if $S+A$ is
completely controllable, $e^{\cal L}=SU(n_Sn_A)$ in particular
contains every matrix of the form $T_A \otimes {\bf 1}$, with $T_A
\in SU(n_S)$, and the claim follows from (\ref{propparttra}) using
$\rho_{TOT}:=\rho_S \otimes \rho_A$.

The rest of the paper is devoted to proving the other direction of
Theorem \ref{MainT}. In section \ref{preres} we give some
preliminary technical results after which, the proof is presented in
section \ref{proofM}. We give some concluding remarks  in section
\ref{conclus}.


\section{Preliminary Results}\label{preres}

\bl{Lemma1} Consider two matrices $X$ and $Y$ in $su(n)$. Then $X$
and $Y$ are linearly dependent if and only if $[X,Y]=0$ and,  for
every $A \in su(n)$,  \be{CC1} \left[ [A,X],\, [A,Y] \right]=0. \ee
\el

\bpr One direction is straightforward. If $X$ and $Y$ are linearly
dependent, then we can write $X=\alpha Y$ (or $Y=\alpha X$) , for
some $\alpha \in \RR$. Then we have $[X,Y]=[\alpha Y,Y]=\alpha
[Y,Y]=0$. Furthermore,  for arbitrary $A \in su(n)$, we have
 \be{CC2} \left[ [A,X],\, [A,Y] \right]=
 \left[ [A,\alpha Y],\, [A,Y] \right]=
 \alpha \left[ [A,Y],\, [A,Y] \right]=0.
 \ee

To prove the converse implication, we first notice that since $X$
and $Y$ commute, they can be simultaneously diagonalized. By
applying the same similarity transformation to all elements in
$su(n)$, there is no loss of generality in assuming that $X$ and $Y$
are both diagonal. Moreover this proves the Lemma for $n=2$, since
we can write $X$ as $X=\alpha \sigma_z$, and $Y$ as $Y=\beta
\sigma_z$, for some real numbers $\alpha$ and $\beta$ and $\sigma_z$
denoting the Pauli
$z-$matrix,\footnote{$\sigma_z:=\left(\begin{matrix}i & 0 \cr 0 &
-i\end{matrix} \right)$} which gives $\alpha Y - \beta X=0$.
Therefore, we can assume $n \geq 3$. Let us denote by $A_{jk}$, with
$j \not=  k$, the  matrix in $su(n)$ \be{CC3} A_{jk}:=|j \rangle
\langle k|- |k\rangle \langle j|. \ee  For $j\not=k$, let us also
denote by $E_{jk}$ the matrix $E_{jk}:= i |j \rangle \langle k|+i |k
\rangle \langle j|$. By writing $X:=\sum_{l=1}^n ix_l |l \rangle
\langle l|$, a straightforward calculation shows that \be{CC4}
[A_{jk}, X]=(x_k-x_j)E_{jk}:=X_{kj}E_{jk},  \ee where we used the
definition $X_{kj}:=x_k-x_j$. Also, using the definition
$Y_{kj}:=y_k-y_j$, we have $[A_{jk},Y]=Y_{kj} E_{jk}$. Now, with
these notations, fix two indices $a$ and $b$ in $\{1,2,\ldots,n\}$,
with $a \not= b$. Consider another index $g$ in $\{1,2,\ldots,n\}$
different from both $a$ and $b$.\footnote{It exists since $n \geq
3$.}  Consider $A=A_{ab}+A_{ga}$. We have from (\ref{CC1}) and
(\ref{CC4}), \be{CC51} 0=\left[ [A,X],[A,Y]\right]= [X_{ba}
E_{ab}+X_{ag} E_{ga}, Y_{ba} E_{ab}+Y_{ag}
E_{ga}]=\left(X_{ba}Y_{ag}-X_{ag} Y_{ba} \right) [E_{ab}, E_{ga}],
\ee which implies \be{CC52} X_{ba}Y_{ag}= X_{ag} Y_{ba}. \ee By
choosing $A=A_{ab}+A_{gb}$, we find analogously \be{CC53}
X_{ba}Y_{bg}=X_{bg}Y_{ba}.  \ee  Summing (\ref{CC52}) and
(\ref{CC53}), we find for any $a$, $b$ and $g$,\footnote{The
equation is obvious  for $g=a$ or $g=b$ or $a=b$.}  \be{CC54}
X_{ba}(y_a+y_b-2y_g)=Y_{ba}(x_a+x_b-2x_g). \ee

Summing  the equations (\ref{CC54}) over all $g$ different from $a$
and $b$. We obtain \be{CCF1}
X_{ba}\left((n-2)(y_a+y_b)-2\sum_{g\not=a, \, g\not=b} y_g \right) =
Y_{ba}\left((n-2)(x_a+x_b)-2\sum_{g\not=a, \, g\not=b} x_g \right).
\ee Using the fact that both $X$ and $Y$ have zero trace we can
replace $\sum_{g\not=a, \, g\not=b} y_g$, with $-(y_a+y_b)$ and
$\sum_{g\not=a, \, g\not=b} x_g$ with $-(x_a+x_b)$ in the above
equation. Recalling the definition of $X_{ba}$ and $Y_{ba}$, we have
$(x_b-x_a)(y_a+y_b)=(y_b-y_a)(x_a+x_b),$ which gives \be{CCAT}
x_by_a=x_ay_b.   \ee This equation is valid for any pair $a$ and $b$
in $\{1,2,\ldots,n\}$. Equation (\ref{CCAT}) is equivalent to $X$
and $Y$ being linearly dependent.\footnote{In fact, if $X=\alpha Y$
(or $Y=\alpha X$) for some real number $\alpha$,  equations
(\ref{CCAT}) are automatically satisfied. Viceversa, assume
(\ref{CCAT}) are verified. If at least one between $X$ and $Y$ is
zero, then they are clearly linearly dependent. Assume that they are
both nonzero and let $\bar a$ be the smallest index $a$ so that at
least one between $x_{a}$ and $y_a$ is different from zero. If
$x_{\bar a}\not=0$ then from (\ref{CCAT}) with $a=\bar a$ we have
that if $y_{\bar a}=0$ then $y_b=0$ for any other $b$ which implies
$Y=0$ which we have excluded. Therefore $y_{\bar a}$ is also
different from zero. For all $b
>\bar a$,  $\frac{x_b}{x_{\bar a}}=\frac{y_b}{y_{\bar a}}$.}

\epr

\bl{Lemma2} (Simplicity Lemma) Consider an element $X \in su(n)$
different from zero  and the space ${\cal V}$ defined
as\footnote{For a general subspace ${\cal P}$ of $u(n)$, and a Lie
subalgebra ${\cal L}$ of $u(n)$, the spaces $ad_{{\cal L}}^k {\cal
P}$ are defined recursively as $ad_{{\cal L}}^0 {\cal P}:= {\cal
P}$, $ad_{{\cal L}}^k {\cal P}:= [{\cal L},ad_{{\cal L}}^{k-1} {\cal
P} ] $.}\be{defiV} {\cal V}:=\bigoplus_{k=0}^\infty \, ad_{su(n)}^k
\, \texttt{span} \{ X \}. \ee Then ${\cal V}=su(n)$. \el \bpr The
space ${\cal V}$ defined in (\ref{defiV}) is an ideal in $su(n)$ and
it is nonzero since $X\not=0$. Since $su(n)$ is simple it has no
nontrivial ideals. So it must be ${\cal V}=su(n)$. \epr

\bl{Lemma3} (Disintegration Lemma)  Consider a  matrix of the form
of $J$ as in (\ref{Jform}) \be{interactionmat} J:=K\otimes {\bf 1} +
{\bf 1} \otimes L+ \sum_{j=1}^n iS_j \otimes \sigma_j,   \ee $K$ and
$L$, matrices in $su(n_S)$ and $su(n_A)$, respectively, and with
$\sigma_j$ linearly independent matrices in $su(n_A)$ and $S_j$
general non zero matrices in $su(n_S)$. The Lie algebra, ${\cal
L}_1$, generated by $\tilde {\cal B}:=\{ {\bf 1} \otimes \sigma \, |
\, \sigma \in su(n_A) \}$ and $J$, is the same as the Lie algebra,
${\cal L}_2$, generated by $iS_1 \otimes \sigma _1$,...,$iS_n
\otimes \sigma_n$, $K \otimes {\bf 1}$  and $ \tilde {\cal B}:=\{
{\bf 1} \otimes \sigma \, | \, \sigma \in su(n_A) \}$. \el

\bpr The inclusion ${\cal L}_1 \subseteq {\cal L}_2$ is obvious
since $J$ is a linear combination of $K \otimes {\bf 1}$, $iS_1
\otimes \sigma _1,$...,$iS_n \otimes \sigma_n$ and an element of
$\tilde {\cal B}$. For the other inclusion, since ${\bf 1} \otimes
L$ is in ${\tilde {\cal B}}$,  ${\cal L}_1$ is generated by $\tilde
{\cal B}$ and \be{formaJ1} J':=K \otimes {\bf 1} + \sum_{j=1}^n iS_j
\otimes \sigma_j.  \ee Then we show by induction on $n$ that $K
\otimes {\bf 1}$ and $i S_1 \otimes  \sigma_1,\ldots, iS_n \otimes
\sigma_n$ are in ${\cal L}_1$. For $n=0$, this is obvious and for
$n=1$, take the Lie bracket of $J'$ with ${\bf 1} \otimes T$, for
some $T$ in $su(n_A)$ so that $[\sigma_1,T] \not=0$. Then $iS_1
\otimes [\sigma_1,T]$ is in ${\cal L}_1$ and, by the simplicity
Lemma \ref{Lemma2}, $iS_1 \otimes \sigma_1$ is in ${\cal L}_1$ so
that $K \otimes {\bf 1}$ is  in ${\cal L}_1$ as well. Assume now $n
\geq 2$. There are two cases to be treated separately. In the first
case, there exists at least one pair $\{\sigma_j, \sigma_k\}$ in
$\{\sigma_1, \sigma_2, \ldots, \sigma_n\}$ such that $[\sigma_j,
\sigma_k] \not=0$. In the second case, all the elements, $\sigma_1,
\sigma_2, \ldots, \sigma_n$, commute.

{\it{Case 1:}} Assume, without loss of generality, that $\sigma_1$
does not commute with all the remaining $\sigma_2,\ldots, \sigma_n$.
Also assume, without loss of generality, that the first $r-1 >0 $
commutators  $[\sigma_2, \sigma_1]$,...,$[\sigma_r,\sigma_1]$ form a
linearly independent set while the remaining $n-r$ commutators (if
any), $[\sigma_{r+1}, \sigma_1]$,...,$[\sigma_n, \sigma_1]$,  can be
each written as linear combinations of the first $r-1$ ones.
Therefore, we write \be{lincomb} [J', {\bf 1} \otimes
\sigma_1]:=\sum_{j=2}^{r} i S_j \otimes [\sigma_j, \sigma_1] +
\sum_{j=r+1}^n i S_j \otimes [\sigma_j, \sigma_1],  \ee and, for
$j=r+1,\ldots,n$, \be{ppps} [\sigma_j,\sigma_1]:=\sum_{l=2}^r a_j^l
[\sigma_l,\sigma_1],  \ee for some coefficients $a_j^l$,
$j=r+1,\ldots, n$, $l=2,\ldots,r$. Defining, for $j=r+1, \ldots, n$,
\be{defXj} X_j:=\sigma_j-\sum_{l=2}^r a_j^l \sigma_l,\ee we notice
that, from (\ref{ppps}), all $X_j$'s commute with $\sigma_1$.
Moreover $\{ \sigma_1,\sigma_2,\ldots,\sigma_r,X_{r+1},\ldots,X_n
\}$ form a linearly independent set. By replacing $\sigma_j$ with
$X_j + \sum_{l=2}^r a_j^l \sigma_l$ using (\ref{defXj}), we can
write $J'$ as \be{newHI} J':=K \otimes {\bf 1}+i S_1 \otimes
\sigma_1 + \sum_{l=2}^r i(S_l +\sum_{j=r+1}^n a_j^l S_j) \otimes
\sigma_l + \sum_{l=r+1}^n iS_l \otimes X_l. \ee Now, if one of the
$(S_l +\sum_{j=r+1}^n a_j^l S_j)$'s, for $l=2,3,\ldots,r$, is zero,
the claim follows by induction on $n$. More precisely, it follows by
induction on $n$, that $iS_1 \otimes \sigma_1$ belongs to ${\cal
L}_1$. Applying the inductive assumption  to $J'-iS_1 \otimes
\sigma_1$, we obtain that $K \otimes {\bf 1}$ and all the matrices
$iS_j \otimes \sigma_j$, $j=2,\ldots,n$, also belong to ${\cal
L}_1$. If all the matrices $(S_l +\sum_{j=r+1}^n a_j^l S_j)$, for
$l=2,3,\ldots,r$, are different from zero, with the expression
(\ref{newHI}) of $J'$ and using the fact that the $X_j$'s commute
with $\sigma_1$, we calculate \be{calcu} [J', {\bf 1} \otimes
\sigma_1]=\sum_{l=2}^r i\left(S_l +\sum_{j=r+1}^n a_j^l S_j\right)
\otimes [\sigma_l,\sigma_1], \ee and since all
$[\sigma_l,\sigma_1]$, $l=2,\ldots,r$, are linearly independent, it
follows from induction that all matrices $i(S_l +\sum_{j=r+1}^n
a_j^l S_j) \otimes [\sigma_l,\sigma_1]$, $l=2,\ldots,r$, belong to
the Lie algebra ${\cal L}_1$. Moreover by taking repeated Lie
brackets with elements ${\bf 1} \otimes \sigma$, with arbitrary
$\sigma \in su(n_A)$,  and taking linear combinations,  it follows
from the simplicity Lemma \ref{Lemma2} that every matrix $i(S_l
+\sum_{j=r+1}^n a_j^l S_j) \otimes \sigma_l $, $l=2,\ldots,r$, also
belongs to ${\cal L}_1$. Therefore these matrices can be subtracted
from $J'$ in (\ref{newHI}) and the claim follows again by induction
on $n$.

\vs

{\it{Case 2}:} The proof is similar to the one of Case 1 but with
some extra complications  due to the fact that all the $\sigma_j$,
$j=1,\ldots n$, commute. Again we use induction on $n$. Given the
form of $ J'$ in (\ref{formaJ1}) and the fact that, in particular,
$\sigma_1$ and $\sigma_2$ are linearly independent, it follows from
Lemma \ref{Lemma1} that there must exist a matrix $A$ in $su(n_A)$
such that $[[\sigma_2,A], [\sigma_1,A]] \not = 0$. By calculating
$\tilde J':=[[J',{\bf 1} \otimes A],{\bf 1} \otimes [\sigma_1,A]]$,
we see that ${\cal L}_1$  contains the matrix \be{HItilde} \tilde
J':=\sum_{j=2}^n i S_j \otimes [[\sigma_j, A], [\sigma_1,A]]. \ee
Let $m$ be the largest integer $(\leq n)$ such that all
$[[\sigma_j,A], [\sigma_1,A]]$, for $j=2,3,\ldots,m$, are linearly
independent. Notice $m$ is at least $2$ because $[[\sigma_2,A],
[\sigma_1,A]] \not = 0$. Therefore we can write $\tilde J'$ in
(\ref{HItilde}) as \be{newtildeHI} \tilde J'=\sum_{j=2}^m i S_j
\otimes [[\sigma_j, A], [\sigma_1,A]]+ \sum_{j=m+1}^n i S_j \otimes
[[\sigma_j, A], [\sigma_1,A]],  \ee with, for every
$j=m+1,\ldots,n$, \be{lineardip2} [[\sigma_j,A], [\sigma_1,A]]:=
\sum_{k=2}^m \alpha_{j}^k [[\sigma_k,A],[\sigma_1,A]], \ee for some
coefficients $\alpha_j^k$, $j=m+1,\ldots,n$, $k=2,\ldots,m$.
Defining, for $j=m+1,\ldots,n$, \be{Xjnew} X_j:= \sigma_j
-\sum_{k=2}^m \alpha_j^k \sigma_k, \ee we have that
$\{\sigma_1,\ldots, \sigma_m,X_{m+1},\ldots,X_n\}$ are linearly
independent and, using (\ref{lineardip2}), \be{commusero}
[[X_j,A],[\sigma_1,A]]=0.  \ee With this definition, $J'$ in
(\ref{formaJ1})  can be written as \be{newHInew} J'=K \otimes {\bf
1} + i S_1 \otimes \sigma_1 + i \sum_{k=2}^m \left( S_k +
\sum_{j=m+1}^n \alpha_{j}^k S_j \right) \otimes
 \sigma_k + \sum_{j=m+1}^n i S_j \otimes X_j.  \ee
If one of the $\left( S_k + \sum_{j=m+1}^n \alpha_{j}^k S_j
\right)$'s  is zero then the claim follows by induction on $n$. In
fact it follows by induction that $iS_1 \otimes \sigma_1$ is in
${\cal L}_1$ and subtracting this from $J'$ in (\ref{formaJ1}), we
can apply the inductive assumption on $n$. If all of these matrices
are different from zero, we consider again $\tilde J'$ in
(\ref{newtildeHI}) calculated by taking the commutator of $J'$ in
(\ref{newHInew}) with ${\bf 1} \otimes A$ and then with ${\bf 1}
\otimes [\sigma_1,A]$ and using (\ref{commusero}). We have \be{tHI}
\tilde J':= \sum_{k=2}^m i\left( S_k + \sum_{j=m+1}^n \alpha_{j}^k
S_j \right) \otimes [[\sigma_k,A], [\sigma_1,A]], \ee which, by the
inductive assumption, gives that all $i\left( S_k + \sum_{j=m+1}^n
\alpha_{j}^k S_j \right) \otimes [[\sigma_k,A], [\sigma_1,A]]$,
$k=2,\ldots,m$ are in ${\cal L}_1$. By the simplicity Lemma
\ref{Lemma2} all $i\left( S_k + \sum_{j=m+1}^n \alpha_{j}^k S_j
\right) \otimes \sigma_k$, $k=2,\ldots,m$, are also in ${\cal L}_1$.
Subtracting them all from (\ref{newHInew}) and applying again the
inductive assumption, we find that $iS_1 \otimes \sigma_1$ is in
${\cal L}_1$, which subtracted from (\ref{formaJ1}) and applying the
inductive assumption once again says that all of the $iS_j \otimes
\sigma_j$, $j=1,\ldots,n$ as well as $K \otimes {\bf 1}$ are in
${\cal L}_1$. This concludes the proof of the Lemma. \epr



\section{Proof of Theorem \ref{MainT}}\label{proofM}

We shall use the following general criterion of indirect
controllability which was proved in \cite{IoeRaf}. Let $\rho_S
\otimes \rho_A$ be the initial state of the system $S+A$ and ${\cal
L}$ the dynamical Lie algebra associated with the dynamics of $S+A$.
Define the subspace of $u(n_S n_A)$, \be{observaspa} {\cal V}:=
\bigoplus_{k=0}^\infty ad_{\cal L}^k \left( \texttt{span}\{i \rho_S
\otimes \rho_A \}\right). \ee Then we have the following theorem
\cite{IoeRaf}. \bt{fromIoeRaf} Let $\rho_S \not=\frac{1}{n_S}{\bf
1}_{n_S}$ and assume that for all $X \in SU(n_S)$ there exists $U
\in e^{\cal L}$ such that \be{tracc} Tr_A (U \rho_S \otimes \rho_A
U^\dagger) = X\rho_S X^\dagger. \ee Then \be{mainsatisf} Tr_A({\cal
V})=u(n_S). \ee \et As a corollary, recalling the Definitions
\ref{indcongivenstate} and \ref{stronglyinUnit}, we have:
\bc{Corollario} Assume that the system $S$ is indirectly
controllable given $\rho_A$,  then the dynamical Lie algebra ${\cal
L}$ is such that, for every $n_S \times n_S$ density matrix of $S$,
$\rho_S \not= \frac{1}{n_S} {\bf 1}_{n_S}$, ${\cal V}$ in
(\ref{observaspa}) satisfies (\ref{mainsatisf}). In particular, if
$S$ is strongly indirectly controllable, then (\ref{mainsatisf}) is
satisfied for every $\rho_A$.  \ec

\vs

We are now ready to prove Theorem \ref{MainT}.

\vs

\bpr Let ${\cal B}_A$ be an orthogonal  basis of $su(n_A)$,   ${\cal
B}_A:=\{ \sigma_1,\ldots,\sigma_{d_A}\}$, where $d_A:=n_A^2-1$ is
the dimension of $su(n_A)$. Every element of ${\cal L}$ can be
written as $J$ in (\ref{Jform}), and using Lemma \ref{Lemma3}, a
basis for ${\cal L}$ can be taken of the form

\be{basisforL} {\cal B}_{\cal L}:=\{{\bf 1} \otimes \sigma_1,\ldots,
{\bf 1} \otimes \sigma_{d_A}, \ee
$$
iL_1^1 \otimes \sigma_1, \ldots, iL_{r_1}^1 \otimes \sigma_1,
$$
$$
iL_1^2 \otimes \sigma_2, \ldots, iL_{r_2}^2 \otimes \sigma_2,
$$
$$
\vdots
$$
$$
iL_1^{d_A} \otimes \sigma_{d_A}, \ldots, iL_{r_{d_A}}^{d_A} \otimes
\sigma_{d_A},
$$
$$
D_1 \otimes {\bf 1}, \ldots, D_s \otimes {\bf 1}\},
$$
where, for every $j=1, \ldots, d_A$, $L_1^j,\ldots,L_{r_j}^j$ can be
taken orthogonal matrices in $su(n_S)$.
Also
$\{D_1,\ldots,D_s\}$ are linearly independent matrices in $su(n_S)$.
To see this  in more detail, notice that every element of ${\cal L}$
can be written as $J$ in (\ref{Jform}), where the $S_j$'s are
orthogonal matrices in $su(n_S)$ and the $\sigma_j$ are orthogonal
matrices in $su(n_A)$ (belonging to a previously chosen basis
$\{\sigma_1,\ldots,\sigma_{d_A}\}$). This is true in particular for
the elements of a given basis of ${\cal L}$. Applying Lemma
\ref{Lemma3}, every element in this  basis can be broken into single
tensor products and all the tensor products so obtained form a
spanning set for ${\cal L}$. Select, in this set, a maximum number
of linearly independent elements. There will be elements of the form
${\bf 1} \otimes F_1,\ldots,{\bf 1}\otimes F_{d_A}$, with
$F_1,\ldots,F_{d_A} \in su(n_A)$  which can be replaced by the
elements as in the first line of (\ref{basisforL}), as well as the
other elements in (\ref{basisforL}).
In summary: It follows from Lemma \ref{Lemma3} that a basis of
${\cal L}$ can be taken
made up of tensor product matrices.

Let $d_S:=n_S^2-1$ be  the dimension of $su(n_S)$.There are three
possible cases:

\begin{enumerate}

\item $\{D_1,D_2,\ldots,D_s\}$ span $su(n_S)$, i.e., $s=d_S$.

\item $s=0$.

\item (intermediate case) $1 \leq s < d_S$.

\end{enumerate}

\vs

In the first case, since there is at least one element in the basis
of ${\cal L}$ of the form $iB \otimes C$ with $B \in su(n_S)$ and $C
\in su(n_A)$, both different from zero,\footnote{This is because the
interaction term in (\ref{Jform}) is assumed different from zero.}
it follows from the simplicity Lemma \ref{Lemma2} applied to both
the $S$ and the $A$ part of the tensor product the {\it all} tensor
product matrices of the form $iB\otimes C$ are in ${\cal L}$ and
therefore ${\cal L}=su(n_Sn_A)$ and $S+A$ is completely
controllable. To conclude the proof of the theorem, we have to show
that, under the indirect controllability (given
$\rho_A=\frac{1}{n_A} {\bf 1}$)  assumption, the other two cases are
not possible.


Consider the second case. To see that it is not possible, notice
that there are, in the basis of ${\cal L}$, at least two matrices
$iA \otimes \sigma_1$ and $iB\otimes \sigma_2$ with $A$ and $B$ in
$su(n_S)$ {\it non-commuting} and some $\sigma_1$ and $\sigma_2$
matrices in $su(n_A)$. If this was not the case, we could choose
($\rho_A=\frac{1}{n_A} {\bf 1}_{n_A}$ and) $\rho_S$ commuting with
all the matrices in the left hand side of the tensor products in the
basis of ${\cal L}$. With this choice, $\rho_S \otimes \rho_A$
commutes with ${\cal L}$ and with all elements in $e^{\cal L}$ and
therefore indirect controllability is not verified ($\rho_S$ is a
fixed point of the dynamics (\ref{combinedevo})). From  the fact
that the matrices $iA \otimes \sigma_1$ and $iB \otimes \sigma_2$
(with $A$ and $B$ non commuting) are  in ${\cal L}$, using the
simplicity Lemma \ref{Lemma2}, it follows that every matrix of the
form $iA \otimes \sigma$ and $iB \otimes \sigma$ with arbitrary
$\sigma \in su(n_A)$ also belongs to ${\cal L}$. Assume $n_A$ is
even and let $\sigma_e$ be the  matrix with alternating $1$ and $-1$
on the diagonal and zero everywhere else (so that the trace is equal
to zero). By calculating $[A \otimes \sigma_e, B \otimes \sigma_e]$,
using the formula $[A \otimes C, B \otimes D]=\frac{1}{2}\left( \{
A,B \} \otimes [C,D]+ [A,B] \otimes \{ C,D \}
\right)$,\footnote{$\{A,B\}$ here denotes the {\it anticommutator},
$\{A,B\}:=AB+BA$.} we obtain \be{ghg} [A \otimes \sigma_e, B \otimes
\sigma_e]=[A,B] \otimes {\bf 1}_{n_A}, \ee which, since $A$ and $B$
do not commute, contradicts our assumption on the basis of ${\cal
L}$. In the case where $n_A$ is odd,  let $\sigma_o^j$, be the
diagonal matrix having alternating $+1$ and $-1$ on the main
diagonal, except in the position $j$ which is occupied by $0$ (so
that $Tr(\sigma_o^j)=0$) and zeros everywhere else. As before, we
calculate \be{ghg1} \frac{1}{n_A -1} \sum_{j=1}^{n_A} [A \otimes
\sigma_o^j, B \otimes \sigma_o^j]=[A,B] \otimes {\bf 1}, \ee which
also contradicts the assumption on the basis of ${\cal L}.$

\vs

The third case is also not possible. To see this,  choose
$\rho_S:=\frac{1}{n_S}{\bf 1}_{n_S}-\alpha i D_1$ with $|\alpha|$
different from zero  but small enough so that $\rho_S$ is still
positive semi-definite. With  $\rho_A=\frac{1}{n_A}{\bf 1}_{n_A}$,
it follows from an inductive argument  that ${\cal V}$ in
(\ref{observaspa}) satisfies \be{fffg} {\cal V} \,  \subseteq \,
{\cal L} \oplus \texttt{span} \{ i {\bf 1}_{n_S} \otimes {\bf
1}_{n_A} \}. \ee Taking the partial trace of both sides in
(\ref{fffg}), we have that \be{hhiq} Tr_A({\cal V}) \, \subseteq \,
\texttt {span} \{ D_1,\ldots,D_s\} \, \oplus \, \texttt{span}
\{i{\bf 1}_{n_S}\}, \ee which since $s < d_S$ contradicts Theorem
\ref{fromIoeRaf}. This concludes the proof of the theorem.

\epr

\section{Concluding Remarks }\label{conclus}

I have proved that indirect controllability and complete
controllability are equivalent notions under appropriate
assumptions. This  extended the equivalence result proved in
\cite{IoeRaf} (Theorem 4) from the case of two qubits to the case of
target system $S$ and accessor system $A$ of arbitrary dimensions.
The result in \cite{IoeRaf} was proven by listing the various
possibilities for the dynamical Lie algebra ${\cal L}$. This list
also showed that, if we choose the initial state of the accessor,
$\rho_A$, as a {\it pure state}, it is not necessary that ${\cal L}$
is the full Lie algebra $su(n_S n_A)$ in order to have indirect
controllability  on $S$. In fact a Lie algebra isomorphic to the
symplectic Lie algebra $sp(2)\,$\footnote{Recall that $sp(n)$ is the
Lie algebra of skew-Hermitian $2n\times 2n$ matrices $A$ satisfying
$JA+A^TJ=0, $ where $J=\left(\begin{matrix}0 & {\bf 1}_n \cr - {\bf
1}_n & 0
\end{matrix} \right)$.} is possible and induces arbitrary  unitary
state transfers for the target system $S$ (Proposition 5.2 in
\cite{IoeRaf}). Our result here was proved using the fully mixed,
maximum entropy, state for the accessor $A$ (as opposed to a pure
state). It is therefore reasonable to expect that, in general, the
`size' of the dynamical Lie algebra ${\cal L}$ needed in order to
have controllability on the state of $S$ will depend on the
eigenvalues of  $\rho_A$, and this dependence is currently under
study.

The two main assumptions of this paper  have been 1) that the
initial state $\rho_{TOT}$ of the system $S+A$ is a product state,
i.e., it is of the form $\rho_S \otimes \rho_A$ and 2) we have full
control on the system $A$. The first assumption corresponds to
starting an experiment with the two system $S$ and $A$ uncorrelated.
From a theory point of view, separating $\rho_S$ and $\rho_A$ in the
initial condition, allowed us to separate the role of $S$ and $A$ in
the definition of indirect controllability and state it as a
property of $S$ only given the set-up for $A$. If the initial state
$\rho_{TOT}$ of $S+A$ is not a product state, we can still define
indirect controllability by requiring that for every $X \in SU(n_S)$
there exists a $U\in e^{\cal L}$ such that $Tr_A(U\rho_{TOT}
U^\dagger)=X Tr_A(\rho_{TOT}) X^\dagger$ for any possible value of
$Tr_S(\rho_{TOT})$. However, there are many $\rho_{TOT}$ giving the
same value of $Tr_S(\rho_{TOT})$, and one should decide how to
restrict in a physical meaningful way the set of such
$\rho_{TOT}$'s. In any case, since much of the machinery developed
in this paper, and in particular the technical results of section
\ref{preres}, dealt with properties of the Lie algebra ${\cal L}$,
The results presented here  can be used to analyze cases where the
initial state  of $S+A$ is not a product state. Even Theorem
\ref{fromIoeRaf} which was proved in \cite{IoeRaf} can be extended
to this case with only notational modifications. The assumption 2)
is used in the technical results of section \ref{preres} and in
particular in the Lemmas \ref{Lemma2} and \ref{Lemma3}. It allowed
us to write the basis of ${\cal L}$ in the convenient form
(\ref{basisforL}) from which we could deduce the main result. If
this assumption is not verified a basis made up of tensor products
might not exist (see, e.g., the examples in section IV-D of
\cite{IoeRaf}). The study of indirect controllability in these cases
will probably  require further analysis and new tools and it remains
an open problem.

\vs

\vs

\vs

{\bf Acknowledgement} This research was supported by NSF under Grant
No. ECCS0824085, and by the ARO MURI grant W911NF-11-1-0268. The
author would like to thank Yao Fang who participated in an
undergraduate research project on the topic of this paper and
provided helpful suggestions.


\begin{thebibliography}{99}

\bibitem{ind1} D. Burgarth, S. Bose, C. Bruder, and V. Giovannetti,
Local controllability of quantum networks, {\it Physical Review A},
79, 060305(R), (2009).


\bibitem{mikobook} D. D'Alessandro, {\it Introduction to Quantum
Control and Dynamics}, CRC-Press, Boca Raton FL, 2007.

\bibitem{mikocool} D. D'Alessandro, Constructive decomposition of the
controllability Lie algebra for quantum systems, {\it IEEE
Transactions on Automatic Control} June 2010, 1416-1421.

\bibitem{IoeRaf} D. D'Alessandro and R. Romano, Indirect
controllability of quantum systems; A study of two interacting
quantum bits, to appear in {\it IEEE Transactions on Automatic
Control}, special issue on Quantum Control.

\bibitem{ind2} H.C. Fu, H. Dong, X. F. Liu, and C.P. Sun, Indirect
control of quantum systems via an accessor: pure coherent control
without system excitation, {\it Journal of Physics A: Mathematical
and Theoretical}, {\bf 42}, 045303, (2009).

\bibitem{JS} V. Jurdjevi\'c and H. Sussmann, Control systems on Lie
groups, {\it Journal of Differential Equations}, 12,  313-329,
(1972).

\bibitem{shapiro} T. Polack, H. Suchowski and D. Tannor,
 Uncontrollable quantum systems: A classifications scheme based on
 Lie subalgebras, {\it Physical Review A}, vol. 79, p. 053403, 2009.

\end{thebibliography}
\end{document}